\documentclass[11pt]{article}

 \newtheorem{theorem}{Theorem}
 
 \newtheorem{example}{Example}
\usepackage{amsfonts}
\usepackage{amsmath}
\usepackage{color}
\usepackage{graphicx}
\usepackage{float}
\usepackage{natbib}
\usepackage[colorlinks,linkcolor=blue,anchorcolor=blue,citecolor=blue]{hyperref}
\usepackage{appendix}

\newcommand{\ba}{\begin{eqnarray}}
\newcommand{\ea}{\end{eqnarray}}
\newcommand{\bas}{\begin{eqnarray*}}
\newcommand{\eas}{\end{eqnarray*}}
\newcommand{\ben}{\begin{enumerate}}
\newcommand{\een}{\end{enumerate}}
\newcommand{\e}{ { \mathbb{E}}}
\newcommand{\var}{ {\mathbb{V}\rm ar }}

 \def\T{{ \mathrm{\scriptscriptstyle \top} }}

\newcommand{\pr}{ {\rm pr}}

\newcommand{\convergeto}{ {\overset{d}{\longrightarrow \; }}}

\usepackage{authblk}
\begin{document}
\date{}
\title{
Score test   for  missing at random or not
}

\author[]{Hairu Wang}
\author[]{Zhiping Lu\thanks{Corresponding author:  zplu@sfs.ecnu.edu.cn}    }
\author[]{Yukun Liu\thanks{Corresponding author:  ykliu@sfs.ecnu.edu.cn}   }

\affil[]{
    KLATASDS-MOE,
 School of Statistics,
  East China Normal University,
    Shanghai 200062, China}

\renewcommand*{\Affilfont}{\small }
\renewcommand\Authands{ and }
\date{}
\maketitle

\begin{abstract}
Missing data are frequently encountered in various disciplines
and   can be divided into three categories:
missing completely at random (MCAR),
missing at random (MAR)
and missing not at random (MNAR).
Valid statistical approaches to missing data
depend crucially on correct  identification of
the underlying missingness mechanism.
Although the  problem of testing whether
this mechanism is MCAR or MAR has been extensively studied,
there has been very little research on testing MAR versus MNAR.
A critical challenge that is faced when dealing with this problem
is the issue of model identification under  MNAR.
In this paper, under a logistic model for the missing probability,
we develop two   score tests for the problem of whether
the missingness mechanism is MAR or MNAR
under a parametric model and a semiparametric location
model on the regression function.
The score tests require  only parameter estimation
under the null MAR assumption,
which completely circumvents the identification issue.
Our simulations and   analysis of   human immunodeficiency virus data
show that the score tests have well-controlled type I errors
and desirable powers.
\end{abstract}


\section{Introduction}
\label{s:intro}

 Missing data are frequently encountered in economic,
medical and  social science disciplines.
Valid statistical inferences for missing data
depend crucially on correct  identification of the underlying
missingness mechanism,
which was  divided  by \citet{r76} into three categories.
The missingness is called   missing at random  (MAR) or ignorable
if it  does  not depend on the missing values themselves conditioning on the observed data,
and it is called missing not at random (MNAR) or nonignorable otherwise.
A degenerate case of MAR is missing completely
at random (MCAR), where the missingness does not depend on
either the observed or the missing data.

Under the  MAR assumption, both the propensity score and
outcome regression models  are nonparametrically identifiable,
and it is therefore  always tractable to conduct valid inferences.
A wide range of statistical approaches have been developed
for MAR data analysis,  including
likelihood-based approaches  \citep{d77, h01, i90},
multiple imputation  \citep{rs86,vs93,r87},
semiparametric methods \citep{z96,r94}
and the inverse probability weighting method \citep{rr83,r94,r95}.
To alleviate the risk of possible misspecifications of
propensity score or outcome regression models,
estimators of double or multiple robustness
have been proposed  and have attracted much attention \citep{s99,ks07,h13,cy14,h14a,h14b,h16a,h16b,cd17,h19}.
For a more comprehensive literature review on  the analysis of MAR data,
see, for example, \cite{lr02,t06,k13} and references therein.

Things become much more challenging when the MAR mechanism is violated
or   data are MNAR.
The foremost challenge  is
 parameter  identifiability: the underlying generating model
is  often not identifiable based on the observed data.
For MNAR data,  even completely parametric models
for the data-generating model  may be
 not identifiable \citep{h79,q02,g82,m15},
let alone semiparametric models \citep{q02,t03,k11,sw16,l20} or
completely nonparametric models  \citep{rr97}.
When no general identification results are available for
MNAR data, the joint distribution of the full data
can only be identified under specific model assumptions.
A popular condition for model identifiability with MNAR data is
the existence of an `instrumental variable' \citep{w14}
or `ancillary variable' \citep{mt15},
which   does not affect the missingness but may affect the conditional
distribution of the response variable.
Quite a few studies of MNAR data under
the instrumental variable condition have been conducted.
\cite{w14} found that the identification issue
for MNAR data can be overcome with the help of
 instrumental variables, and they
then  proposed the use of the generalized method of moments to estimate model parameters.
This approach was extended by \cite{zs15} to generalized linear models
under a parametric propensity score model,
and this was further extended by \cite{sw16} to allow for the semiparametric propensity model of \cite{k11}.
In the presence of an instrumental variable
or ancillary variable,
double robust estimation and semiparametric efficient estimation
under MNAR data have also been investigated
 \citep[see, e.g.,][]{mk16,mt15,a18,l20}.
However,   an instrumental variable
may not be readily available or may not be straightforward to find in practice,
which complicates  the identifiability and inferences of the
existing statistical approaches.
  \cite{tang2018} and \cite{wk21}
provide more comprehensive reviews of
statistical inferences for nonignorable missing-data problems.

These discussions arguably reveal that  methods for handling MAR data
and MNAR data  are totally different:  the former are relatively easy
whereas the latter are much more difficult.
Correctly determining  which mechanism is responsible for data being missing
is crucial to the subsequent development of  valid inference methods.
This raises the hypothesis testing problem of
whether the missingness mechanism is MAR or MNAR.

A relative simple counterpart of  this hypothesis testing problem is
 whether the missingness mechanism is MCAR or MAR.
Many tests for MCAR have been developed in recent decades,
since the MCAR category is at the centre of interest of many behavioural
and social scientists confronted with missing data \citep{ss86}.
\cite{l98} constructed a test by comparing the means of recorded
values of each variable between groups of different missingness patterns.
\cite{cl99} extended Little's test to longitudinal data
by comparing the means of the general estimating equations
across different missingness patterns with zero, with
any departure from zero then indicating rejection of the MCAR hypothesis.
More extensions of Little's idea
to  comparisons of the means, the covariance matrices
and/or the distributions across different missingness patterns
have also been  investigated \citep[see, e.g., ][]{jj10,kb02,ly15}.
Recently, \cite{z19}
proposed   a nonparametric approach for testing MCAR
based on empirical likelihood \citep{o88,o90,o01}.
Their approach also provides  a unified procedure for estimation after the MCAR hypothesis has been rejected.

In contrast to  MCAR,
testing for MAR has not received much attention so far.
The first contribution
in this direction was  the nonparametric test proposed by \cite{b17},
which was based on an integrated squared distance.
Under a generalized linear regression model,
\cite{d20} proposed to test MAR by a quadratic form
of the difference of the estimators of the regression coefficient
under the MAR and MNAR assumptions, respectively.
To the best of our knowledge, these are the only two formal tests for MAR.
They both require the existence of an instrumental variable
to guarantee identifiability,
because their test statistics depend on consistent estimates under MNAR.
However, as  mentioned previously,
the identification of an instrumental variable
is usually not straightforward,
and, even worse, it may not exist.
Also, consistent parameter estimation itself under MNAR  is
rather challenging.

In this paper, we propose two score tests for MAR
under a linear logistic model
when a completely parametric model and a semiparametric location model, respectively,
are imposed on the outcome regression model, respectively.
Compared with the tests proposed by \cite{b17} and \cite{d20},
the new tests have at least three advantages.
The first remarkable advantage is  that no identification
condition  is required under MNAR, which implies
that no instrumental variable is needed.
However, without an instrumental variable, the tests of
\cite{b17} and \cite{d20} may fail to work.
The second advantage is that  the  new tests involve
 much easier calculations, because the underlying unknown parameters
need only  be estimated under MAR. As we have pointed out,  identifiability is not an issue
for MAR data and
parameter estimation has been well studied.
Third,  our simulation results indicate that
the two new tests are  often more powerful than
or at least comparable to that proposed by \cite{d20}.

%

\section{Score test}
\label{s:full}
Let $Y$ denote an outcome  that is  subject to missingness
and let $X$ be a fully observed  covariate vector
 whose first component is $1$.
We denote by $D$   the missingness indicator of the outcome,
with  $D = 1$ if $Y$ is observed and $0$ otherwise.
We wish to test whether the missingness mechanism is MAR or MNAR, namely
$H_0: \pr(D=1|Y, X) = \pr(D=1|X)$.
Suppose that  the missingness probability or the propensity score
follows a linear logistic model
\ba
\label{propensity}
\pr(D=1|X=x, Y=y)
= \pi(x^\T \beta  + \gamma y)
\ea
with  $\pi(t) = e^t/(1+e^t)$.
Under the model \eqref{propensity},
the testing problem of interest is equivalent to
$H_0:  \gamma = 0$, because
the missingness mechanism is MAR if $\gamma = 0$ and  MNAR otherwise.

Suppose that $\{(d_i, d_iy_i, x_i), i=1,2, \ldots, n\}$
are $n$ independent and identically distributed observations from
$(D, DY, X)$.
Let  $f(y|x)$ denote the conditional density function of $Y$ given $X=x$.
The loglikelihood based on the observed data is
\bas
\ell (\gamma, \beta, f)
&=& \sum_{i=1}^n [d_i \{ \log\pi(x_i^\T \beta + \gamma y_i)  +
\log f(y_i|x_i)\} \\
&&
\hspace{2cm}
+ (1-d_i) \log
\int \{ 1-\pi(x_i^\T \beta + \gamma y )\} f(y|x_i )dy].
\eas
The likelihood ratio test  is the most natural
and preferable  for testing $\gamma=0$.
Unfortunately, \cite{m15}  showed that
parameter identifiability
is not guaranteed even when a parametric model is postulated
for  $f(y|x)$.
Without parameter identifiability,
consistent parameter estimation is not feasible, and
therefore nor is  the  likelihood ratio test,
under general parametric assumptions
because these require consistent parameter estimation
under the null and  alternative hypotheses. The
Wald test has the same problem.

The score test was introduced by \cite{r48}  as an alternative to
the likelihood ratio test and Wald test.
The most significant advantage of the score statistic
is that it depends only on estimates of parameters under $H_0$;
in other words, it automatically circumvents the notorious
identifiability issue under MNAR.
This  motivates us to  consider testing  $\gamma=0$ by a score test.
Let $\nabla_{\gamma}$  denote the partial differential operator with respect to $\gamma$.
The score function with respect to $\gamma$ is
\bas
\nabla_{\gamma} \ell (\gamma, \beta, f)  |_{\gamma=0}
=
\sum_{i=1}^n  \left[ d_i   \{ 1-  \pi(x_i^\T \beta ) \} y_i
- (1-d_i)
 \pi(x_i^\T \beta   )  \mu(x_i) \right],
\eas
which depends on the unknown  parameters $\beta$ and $\mu(x) = \int y f(y|x )dy$.

The score test statistic is constructed by replacing $\beta$ and $\mu(\cdot)$
with their estimators  under the null hypothesis $H_0: \gamma = 0$.
The likelihood function under $H_0$ becomes
\bas
\ell_0 ( \beta, f)
&=& \sum_{i=1}^n [d_i   \log\pi(x_i^\T \beta )  + d_i
\log f(y_i|x_i)
+ (1-d_i) \log
  \{ 1-\pi(x_i^\T \beta   )\} ].
\eas
In this situation, a natural estimator for $\beta$ is the  maximum likelihood estimator
$\hat \beta = \arg\max_{\beta} \ell_1(\beta)$, where
$
\ell_1 (\beta )
= \sum_{i=1}^n [d_i \log\pi(x_i^\T \beta  )
+ (1-d_i) \log
  \{ 1-\pi(x_i^\T \beta  )\}
 ]
$
is the likelihood function of $\beta$
under the null hypothesis.
Estimation of $\mu(\cdot)$ depends on model assumptions on $f(y|x) = \pr(Y=y|X=x)$.
To finish the construction of the score test,
we consider postulating either   a fully parametric
or  semiparametric model  on $f(y|x)$.

\subsection{Score test under a parametric model $f(y|x, \xi)$ }

 When $H_0$ or equivalently the missingness mechanism is MAR,
$ f(y|x)=\pr(Y=y|X=x) = \pr(Y=y|X=x, D=1)$.
Under a  fully parametric model $f(y|x, \xi)$ on $f(y|x)$,
this motivates us to estimate $\xi$ by $\hat \xi = \arg\max_{\xi} \ell_2 ( \xi)$, where
$
\ell_2 ( \xi)
=
d_i\log f(y_i|x_i, \xi)
$
is the likelihood function of $\xi$ under $H_0$.
The score test statistic is then
\bas
S_1(\hat \beta, \hat \xi)
=
\sum_{i=1}^n  \left[ d_i   \{ 1-  \pi(x_i^\T \hat\beta ) \} y_i
- (1-d_i)
 \pi(x_i^\T \hat \beta   )   \int y f(y|x_i, \hat \xi )dy\right].
\eas

To calculate the $p$-value of a score test statistic,
we need to determine the sampling distribution of
$S_1(\hat \beta, \hat \xi)$ under $H_0$.
The exact form of this sampling distribution  is in general intractable.
A more practical solution  is to approximate it by
its null limiting distribution under $H_0$ or MAR.

Let $\beta_0$ and $\xi_0$ be the true values of $\beta$ and
$\xi$, respectively.
Our theoretical results on $S_1(\hat \beta, \hat \xi)$ are built on the following regularity conditions
on $X$ and $f(y|x, \xi)$:
\begin{description}
\item[(C1)]
$\e \|X\|^2 < \infty$ and $A = \e[ \pi(X^\T \beta_0 ) \{ 1- \pi(X^\T \beta_0) \}  XX^\T ]$ is of full rank.

\item[(C2)]
(i) The parameter space $\Omega$  of $\xi$  is independent of $(y, x)$  and compact.
(ii) The true value $\xi_0$ of $\xi$ is an interior point of $\Omega$.
(iii)  $\xi$ is identifiable,  i.e.
$\e  \{ \int |f(y|X, \xi) - f(y|X, \xi')| dy \}  >0 $
for any different elements $\xi$ and $ \xi'$ in  $\Omega$.
(iv) $\e\{\sup_{\xi \in \Omega} | \log f(Y|X, \xi)|\}<\infty$.
(v) $f(y|x, \xi)$  is continuous in $\xi$ for almost all $(y, x)$.
\item[(C3)]
(i)  $f(y|x, \xi)$ is twice differentiable with respect to $\xi$ for almost all $(y, x)$,
and  $ \nabla_{\xi \xi }  f(y|x, \xi)$   is continuous at   $\xi_0$.
(ii) $B =   \e[  \pi(X^\T \beta_0 )  \{ \nabla_{\xi  }  \log f(Y|X, \xi_0)\}^{\otimes 2} ]$ is positive definite.
(iii) There exist a $\delta>0$ and
positive functions $M_1(x)$ and $ M_2(y, x)$
 such that   $ \e\{ M_1(X)\} <\infty$ and $ \e \{ M_2(y, x)\}  < \infty$,
and
\bas
 \|x\|  \int |t|\{  f(t|x, \xi)+  \|\nabla_{\xi} f(t|x, \xi)\|\}dt \leq M_1(x) \  \
 \mbox{and} \   \  \|\nabla_{\xi\xi^\T}\log f(y|x, \xi)\|  \leq M_2(y, x)
\eas
for all $\xi$ satisfying $ \|\xi-\xi_0\|\leq \delta$.
\end{description}
Under Condition (C1),
the limit function of $\ell_{1}(\beta)/n$ is well defined.
Conditions (i) and (ii) in Condition (C2) are trivial.
Condition (iii) guarantees that $\xi_0$ is a unique maximizer of the likelihood $\ell_2(\xi)$.
Condition (iv) provides an envelope for  $\{ \log f(y|x, \xi): \xi \in \Omega \}$.
These conditions plus the continuity condition of Condition (C2)(v)
are sufficient for the consistency of $\hat \xi$.
Under Condition (C3), the loglikelihood  $\ell_2(\xi)$ can be approximated
by a quadratic form of $\xi$.
The asymptotic normality of $\hat \xi$ follows  immediately.
Condition (C3) also guarantees that
 the matrices defined in Theorem~\ref{score1-normality} are well defined.

\begin{theorem}\label{score1-normality}
  Assume  Conditions (C1)--(C3) and that  $H_0: \gamma=0$ is true.
As $n$ goes to infinity,
$
n^{-1/2} S_1( \hat\beta,\hat\xi) \convergeto \mathcal{N}(0, \sigma_1^2),
$
where
$\sigma_1^2 = A_2+B_2 - A_1^\T A^{-1}A_1-B_1^\T B^{-1}B_1$, and
\bas
A_2 &=& \e\{  \pi(X^\T \beta_0 )  \{ 1-  \pi(X^\T \beta_0 ) \}^2 Y^2 \},  \\
B_2&=& \e[
  \{ 1-\pi (X^\T \beta_0   )\}
   \{ \pi (X^\T \beta_0   ) \}^2  \{  \int y   f(y|X, \xi_0)dy  \}^2], \\
A_1
&=&
 \e\left[    \pi(X^\T \beta_0 )   \{ 1-  \pi(X^\T \beta_0 ) \} XY \right],
 \\
B_1
&=&
  \e\left[
  \{1-\pi(X^\T \beta_0 ) \}
 \pi(X^\T \beta_0   )   \int y  \nabla_{\xi} f(y|X, \xi_0)dy    \right].
\eas

\end{theorem}

An estimator for the asymptotic variance $\sigma_1^2$ is
$\hat \sigma_1^2
=
\hat A_2+\hat B_2
- \hat A_1^\T \hat A^{-1}\hat A_1
- \hat B_1^\T \hat B^{-1}\hat B_1$,
where
\bas
\hat A
&=&
   \frac{1}{n}\sum_{i=1}^n \pi(x^\T_i \hat \beta ) \{ 1- \pi(x^\T_i \hat  \beta ) \}x_i x_i^\T, \\
\hat B
  &=&
   - \frac{1}{n}\sum_{i=1}^n \pi(x^\T_i \hat  \beta  )
      \int \{ \nabla_{\xi\xi^\T}\log f(y|x_i, \hat \xi )\} f(y|x_i,\hat  \xi)dy,
\\
      \hat B_2
    &=&
     \frac{1}{n}\sum_{i=1}^n \pi^2(x_i^\T  \hat  \beta)
       \{ 1- \pi(x_i^\T \hat  \beta )\} \{\int yf(y|x_i,\hat  \xi)dy\}^2,
  \\
\hat A_1
    &=&
    \frac{1}{n}\sum_{i=1}^n \pi(x^\T_i \hat  \beta )
    \{1-\pi(x^\T_i \hat \beta )\}x_i  \int yf(y|x_i, \hat  \xi )dy, \\
\hat B_1
   &=&
   \frac{1}{n}\sum_{i=1}^n \pi(x^\T_i \hat \beta ) \{1-\pi(x_i^\T\hat \beta )\}
   \int y\nabla_\xi f(y|x_i,\hat \xi)dy,  \\
\hat A_2
  &=&
  \frac{1}{n}\sum_{i=1}^n \pi(x_i^\T \hat \beta )
  \{ 1- \pi(x_i^\T \hat  \beta )\}^2 \int y^2 f(y|x_i,\hat \xi )dy.
\eas
The consistency of $\hat \sigma_1^2$
follows from the consistency of $(\hat \beta, \hat \xi)$
and the continuity of $\pi(x^\T \beta)$ and of $f(y|x, \xi)$.
Formally, the proposed score test rejects $H_0$ if  $ |S_1(\hat \beta, \hat \xi)|/(\sqrt{n}\hat \sigma_1)$
is too large, and its $p$-value is approximately
$2-2\Phi\{  |S_1(\hat \beta, \hat \xi)|/(\sqrt{n}\hat \sigma_1) \}$, where $\Phi(\cdot)$
is the standard normal distribution function.

\subsection{Score test under a semiparametric location model on $f(y|x)$}

 The score function depends on  $f(y|x)$  through
the conditional mean $\mu(x) = \int yf(y|x)dy$, which
 is equal to $\e\{Y|X=x, D=1\}$ under $H_0$.
Instead of imposing a fully parametric conditional density model,
it is sufficient to assume a parametric model $\mu(x, \theta)$  for $\mu(x)$,
where  $\theta$ is an unknown parameter.
Under $H_0$, this model assumption is equivalent to  a location model on the completely observed data
$\{(x_i, y_i): d_i=1\}$,
namely
$ y_i = \mu(x_i,\theta) + \varepsilon_i,$
where $\varepsilon_i $ satisfies
  $\e(\varepsilon_i|X_i=x_i, D_i =1) = 0$.
We estimate   $\theta$ by the least square estimator
\bas
\hat \theta = \arg\min \sum_{i=1}^n  d_i\{ y_i - \mu(x_i, \theta)\}^2.
\eas
Given the estimators $\hat \beta$ and $\mu(x, \hat \theta)$
of $\beta$ and $\mu(x)$,
the score test statistic under the location model on $f(y|x)$ is
\bas
S_2(\hat \beta, \hat \theta)
=
\sum_{i=1}^n  \left[ d_i   \{ 1-  \pi(x_i^\T \hat \beta ) \} y_i
- (1-d_i)
 \pi(x_i^\T \hat \beta   )   \mu(x_i, \hat \theta)\right].
\eas

Let  $\theta_0$ be the true value  of  $\theta$.
To establish the limiting distribution of $S_2(\hat \beta, \hat \theta)$,
we impose the following regularity conditions  $\mu(x, \theta)$:

\begin{description}

\item[(D1)]
(i)  $\e( Y|X ) = \mu(X, \theta_0)$ holds for all $X$.
(ii)  The parameter space $\Theta$ of $\theta$ is compact, and  $\theta_0$ is an interior in $\Theta$.
(iii)    $\theta_0$ is  the unique minimizer of
 $\L_*(\theta) =  \e [ \pi(X^\T \beta_0  )\{Y- \mu(X,\theta)\}^2]$.

\item[(D2)]
(i) $\mu(x,\theta)$ is twice differentiable with respect to $\theta$.
(ii) There exists $ M_3(X, Y)$ such that  $\e\{M_3(X, Y)\}<\infty$
and
$
 Y^2+ \{\mu(X,\theta)\}^2    \leq M_3(X, Y)
$
holds for all $\theta$.
(iii)  There exists $ M_4(X, Y)$ such that $\e\{M_4(X, Y)\}<\infty$ and
\[
 \{ |Y| + |\mu(X, \theta)| \}
 \|\nabla_{\theta\theta^\T} \mu(X, \theta)\|
 + \|\nabla_\theta\mu(X,\theta )\|^2  \leq M_4(X, Y)
\]
holds for all $\theta$.
(iv)  $C_1 = \e[ \{\nabla_\theta \mu(X,\theta_0)\}^{\otimes2} \pi(X^\T \beta_0) ]$
is positive definite and
 $C_2 = \e[\{Y-\mu(X,\theta_0)\}^2 \{\nabla_\theta\mu(X,\theta_0)\}^{\otimes2}
\pi(X^\T \beta_0 )]$ is well defined.

\end{description}
Conditions (D1) and (D2) are the analogues of Conditions (C2) and (C3)
for  the conditional mean model  $\mu(x, \theta)$.

\begin{theorem} \label{score2-normality}
Assume Conditions (C1), (D1) and (D2) and that $H_0:\gamma=0$ is true.
As $n$ goes to infinity,
$
n^{-1/2} S_2( \hat\beta,\hat\theta) \convergeto \mathcal{N}(0, \sigma_2^2),
$
where
$\sigma_2^2 = A_2+B_4 - A_1^\T A^{-1}A_1 + B_3^\T C_1^{-1}C_2C_1^{-1}B_3 - 2B_3^\T C_1^{-1}C_3$ and
\bas
B_3 &=&
  \e\left[
  \{1-\pi(X^\T \beta_0 ) \}
 \pi(X^\T \beta_0   )    \nabla_{\theta}\mu(X,\theta_0)   \right],\\
B_4 &=&
  \e[
  \{ 1-\pi (X^\T \beta_0   )\}
   \{ \pi (X^\T \beta_0   ) \}^2  \{  \mu(X,\theta_0)\}^2], \\
C_1
   &=& \e[ \{\nabla_\theta \mu(X,\theta_0)\}^{\otimes2} \pi(X^\T \beta_0) ],\\
C_2 &=& \e[\{Y-\mu(X,\theta_0)\}^2 \{\nabla_\theta\mu(X,\theta_0)\}^{\otimes2}
\pi(X^\T \beta_0 )], \\
C_3 &=&
  \e [\{ 1-\pi (X^\T \beta_0   )\}  \pi (X^\T \beta_0   )   \{ Y-\mu(X,\theta_0) \}^2 \nabla_\theta\mu(X,\theta_0)].
\eas

\end{theorem}

A consistent estimator for the asymptotic variance $\sigma_2^2$ is
\bas
\hat \sigma_2^2
=
  \hat A_2 + \hat B_4 - \hat A_1^\T \hat A^{-1}\hat A_1
  + \hat B_3^\T \hat C_1^{-1}\hat C_2\hat C_1^{-1}\hat B_3 - 2\hat B_3^\T \hat C_1^{-1}  \hat C_3,
  \eas
where
\begin{eqnarray*}
\hat A
&=&
   \frac{1}{n}\sum_{i=1}^n \pi(x^\T_i \hat \beta ) \{ 1- \pi(x^\T_i \hat  \beta ) \}x_i x_i^\T, \\
\hat A_1
    &=&
    \frac{1}{n}\sum_{i=1}^n \pi(x^\T_i \hat  \beta )
    \{1-\pi(x^\T_i \hat \beta )\}x_i  \mu(x_i,\hat \theta), \\
\hat A_2
  &=&
  \frac{1}{n}\sum_{i=1}^n
  \{ 1- \pi(x_i^\T \hat  \beta )\}^2 [   d_i\{ y_i - \mu(x_i, \hat \theta) \}^2
  +  \pi(x_i^\T \hat  \beta ) \mu^2(x_i,\hat\theta)],
\\
\hat B_3
   &=&
   \frac{1}{n}\sum_{i=1}^n \pi(x^\T_i \hat \beta ) \{1-\pi(x_i^\T\hat \beta )\}
   \nabla_\theta\mu(x_i,\hat \theta),  \\
\hat B_4
    &=&
     \frac{1}{n}\sum_{i=1}^n \pi^2(x_i^\T  \hat  \beta)
       \{ 1- \pi(x_i^\T \hat  \beta )\} \{\mu(x_i,\hat \theta)\}^2,  \\
\hat C_1 &=&
   \frac{1}{n}\sum_{i=1}^n\{\nabla_\theta \mu(x_i,\hat\theta)^{\otimes2} \pi(x_i^\T \hat\beta) \}, \\
\hat C_2 &=&
   \frac{1}{n}\sum_{i=1}^n [   d_i   \{ y_i - \mu(x_i, \hat \theta) \}^2   \nabla_\theta \mu(x,\hat\theta)^{\otimes2}  ],\\
\hat C_3 &=&
  \frac{1}{n}\sum_{i=1}^n [\{ 1-\pi (x_i^\T \hat\beta   )\}
  d_i   \{ y_i - \mu(x_i, \hat \theta) \}^2
  \nabla_\theta\mu(x_i,\hat\theta)].
\end{eqnarray*}
We  reject $H_0$
if  $ |S_2(\hat \beta, \hat \theta)|/(\sqrt{n}\hat \sigma_2)$
is   large enough.
Its $p$-value is approximately
$2-2\Phi\{  |S_2(\hat \beta, \hat \theta)|/(\sqrt{n}\hat \sigma_2) \}$.
The result in Theorem~\ref{score2-normality}
and the variance estimator $\hat \sigma_2^2$ allow
the error $\varepsilon_i$ given $D_i=1$  and $X_i=x$
to depend on $x$, or to have a heterogeneous variance.
If we assume that $x_i$ and $\varepsilon_i$ are
conditionally independent given $D_i=1$,
then
$
C_3 = B_3 \times \var(\varepsilon_i|D_i=1)
$ and
$
C_2 = C_1 \times \var(\varepsilon_i|D_i=1)
$
and $\sigma_2^2$ reduces to
$\sigma_2^2 = A_2+B_4 - A_1^\T A^{-1}A_1 - B_3^\T C_1^{-1}B_3\times \var(\varepsilon_i|D_i=1)$.

\subsection{Local power }

To study the asymptotic power of the proposed score tests, we consider the following local alternative:
\ba\label{alt}
H_a: \gamma=n^{-1/2}\gamma_0,
\ea
where $\gamma_0$ is fixed.
The local alternative \eqref{alt} tends to the null hypothesis at a root-$n$ rate
as $n$ goes to infinity.
A  test for $H_0$ is root-$n$ consistent if
it can detect the local alternative \eqref{alt}
as $n$ goes to infinity for any fixed $\gamma_0$.
We expect that both of
the proposed score tests
have  root-$n$ consistency, which  is a
 desirable property of a nice test for MAR.

\begin{theorem}  \label{score-local-power}
Assume Condition (C1) and that   the alternative $H_a$ is true.
Let $\beta_0, \xi_0$ and $\theta_0$ be the true values of $\beta$, $\xi$
and $\theta$, respectively.
(i) If Conditions (C2) and (C3) are satisfied, then, as $n$ goes to infinity,
$
n^{-1/2} S_1( \hat\beta,\hat\xi) \convergeto \mathcal{N}(\gamma_0 \sigma_1^2, \sigma_1^2),
$
where   $\sigma_1^2 $ is defined in Theorem~\ref{score1-normality}.
(ii) If conditions (D1) and (D2) are satisfied, then,
as $n$ goes to infinity,
$
n^{-1/2} S_2( \hat\beta,\hat\theta) \convergeto \mathcal{N}(\delta, \sigma_2^2),
$
where
$\delta = \gamma_0(A_2+B_4 - A_1^\T A^{-1}A_1-B_3^\T C_1^{-1}C_3)$
and
$\sigma_2^2$ is defined in Theorem~\ref{score2-normality}.
\end{theorem}

Under the propensity model \eqref{propensity},
a nonzero $\gamma$ characterizes the departure of the true missingness mechanism
from the null hypothesis.
Theorem \ref{score-local-power} indicates that
if for some fixed $\gamma_0$ the alternative $H_a$  is true, then
both of the score test statistics  converge  in distribution to
nondegenerate distributions with nonzero location parameters.
Because the absolute  values of the location parameters are increasing functions of   $|\gamma_0|$,
 the powers of both  score tests
 tend to $1$ as $\gamma_0$ goes to infinity, which means that
 both of them are root-$n$ consistent.

\section{Simulation}
\label{s:simulation}
 We conduct simulations to evaluate the finite-sample performance of the proposed score tests.
Specifically, we compare the following three tests:
(1) S1,  the proposed score test under a parametric model on $\pr(y|x)$,
(2) S2,  the proposed score test under a semiparametric
location model on $\pr(y|x)$   and
(3) DUAN,   the test proposed by \cite{d20}.
We generate data from two examples.
In Example~\ref{ex-duan}, which comes from  \cite{d20},
an instrumental variable is present, whereas there is
no instrumental variable  in Example~\ref{ex-no-instrument}.
All our simulation results are calculated based on 5000 simulated samples
and  the significance level is set to 5\%.

\begin{example}\label{ex-duan}
Let $(Y, U, Z)$ follow a multivariate normal distribution
such that   $(Y| U, Z) \sim N(1+U+ b_{z} Z, 1 )$,
$(U \mid Z) \sim N(1-Z, 1)$
and $Z \sim N(0,1)$.
The missingness indicator  $D$ for  $Y$ follows a Bernoulli distribution
with success probability $\pr(D=1 \mid Y, U, Z) \sim \Phi\left(c_{0}+c_{1} w(Y)+c_{2} U\right)$,
where $\Phi(\cdot)$ is  the standard normal distribution function.
We consider three choices for $w(y)$, namely $y, 0.4y^2$ and $ 2.5I(y>1)$,
 two choices for $b_z$, namely 0.5 and 1,
four choices for $c_2$, namely 0, 0.25, 0.5, and 0.75,
 and 11 choices for $c_1$, namely  $0.05\times k$,
 for $k=0, 1, \ldots, 10$.
\end{example}

The DUAN test requires the existence of an instrumental variable.
Under the settings of Example~\ref{ex-duan},
the variable $Z$ is an instrumental variable and
therefore  the DUAN test is applicable.
For data  generated from this example,
we model $\pr(Y=y|X=x)$  by $f(y|x, \xi) = (2\pi)^{-1/2} \exp\{ - (x^\T \xi)^2/2\}$
in the construction of the score test S1,
and we model $\e(Y|X=x)$ by $\mu(x, \theta) = x^\T \theta$ for S2.
Tables~\ref{tab-ex-duan-1000-weak} and
\ref{tab-ex-duan-1000-strong} present the simulated rejection rates
of S1, S2,and  DUAN when the sample size  $n=1000$.
The rejection rates corresponding to the DUAN test are directly
copied from   Tables 3 and 4 in  \cite{d20},
which were calculated based on 1000 simulated samples.
Although the true missingness indicator is generated from a probit model,
we model it by  the logistic model \eqref{propensity}
in the constructions of the proposed two score tests.

The coefficient $c_1$ quantifies the departure of the
 true missingness mechanism from the null hypothesis.
When $c_1 = 0$, the null hypothesis holds and
the results reported are all type I errors.
We see that all three tests have desirable controls on their type I errors.
As $c_1$ increases,  the true missingness mechanism departs more and more
from the null hypothesis
and, as expected, all tests have increasing powers.
The proposed two score tests are  more powerful than DUAN in most situations.
When $b_z=0.5$, their power gains against DUAN can be greater than 25\%; see
the case with $c_1=0.4, c_2=0.75$ and $w(y) = y$.
As $b_z$ increases from 0.5 to 1,
the  power gain  can be as large as  41\%;
see the case with $b_z=1, c_1=0.25, c_2=0.75$ and $w(y)=0.4y^2$.
These observations show that the proposed score tests
have obvious advantages over the DUAN test.
Meanwhile,  the two score tests have almost the same powers in all cases,
although the S2 test requires much weaker model assumptions.
We also conduct simulations for  $n=2000$, and the simulation results,
provided in the Supplementary Material, are similar.

\begin{figure}[htp]
  \centering
  \includegraphics[width=0.45\textwidth]{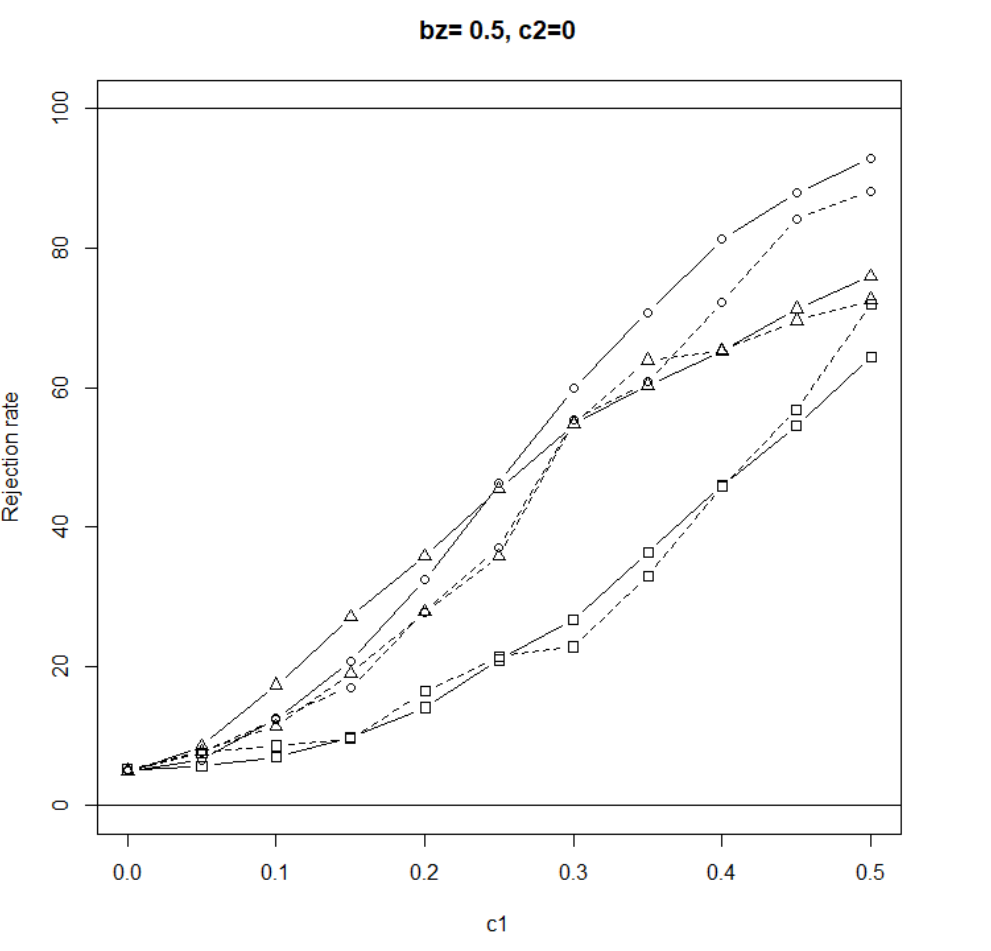}
    \includegraphics[width=0.45\textwidth]{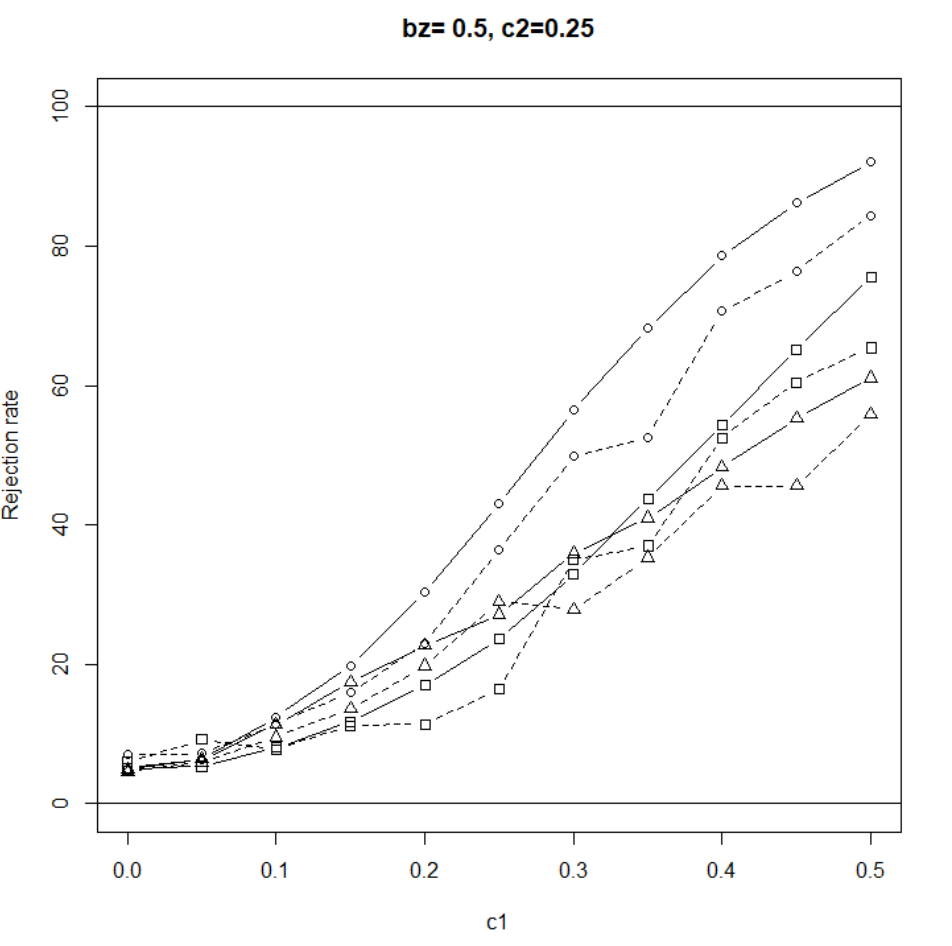}  \\
  \includegraphics[width=0.45\textwidth]{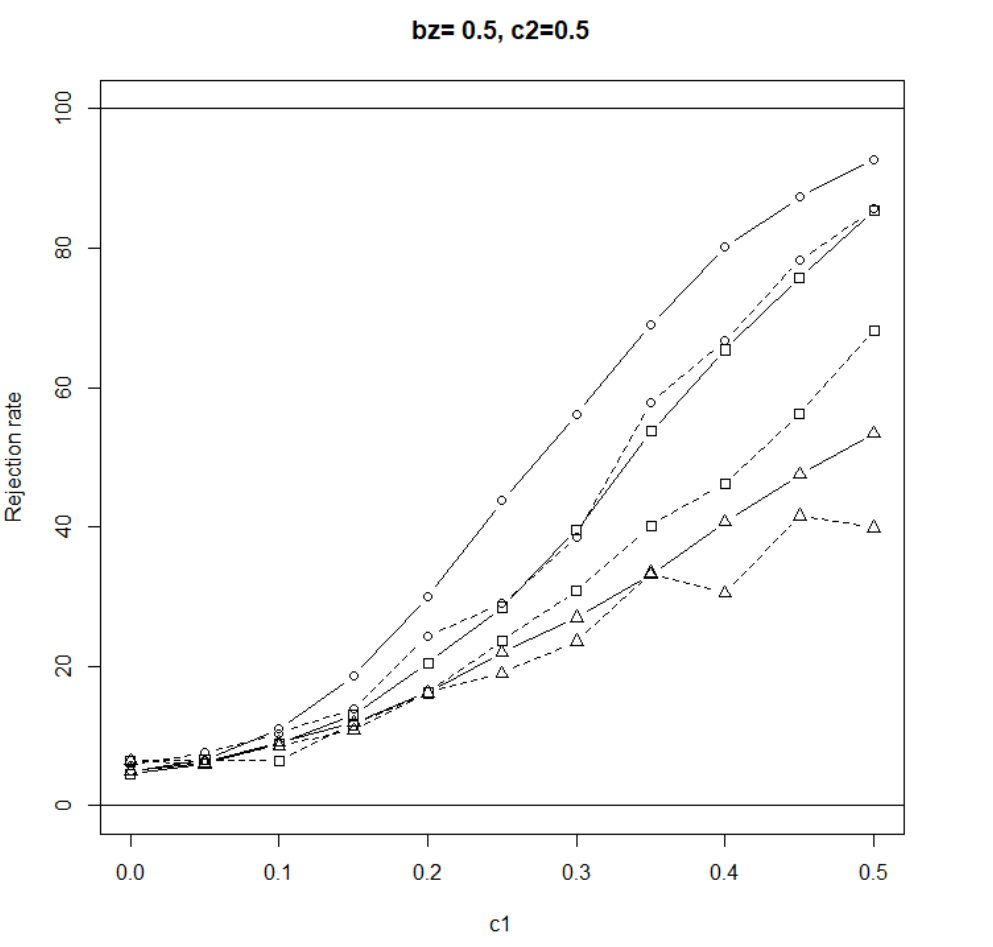}
    \includegraphics[width=0.45\textwidth]{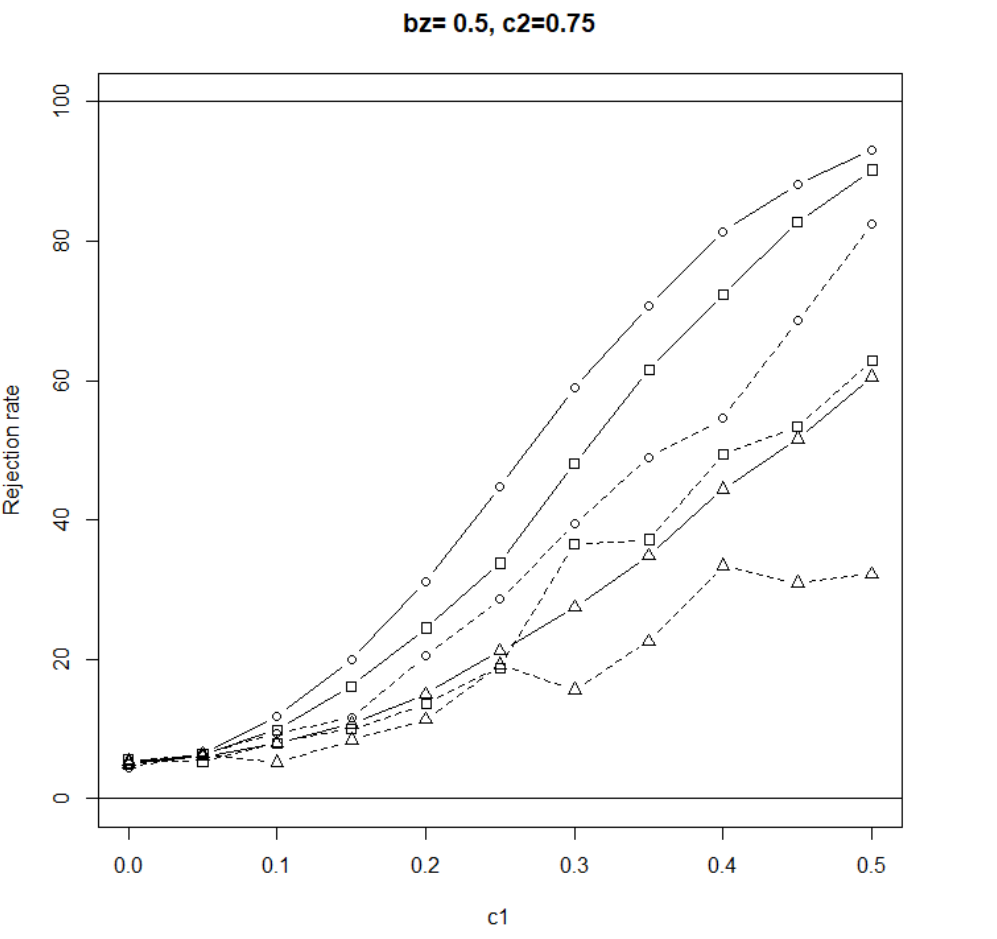}
  \caption{Plots of rejection rates when $b_z=0.5$ for
 the S2 test (solid lines) and the DUAN test  (dotted lines):
$w(y) = y$ (circles);
$w(y) = 0.5y^2$ (triangles);
$w(y) = 2.5 I(y>1)$ (squares).
  }\label{power-bz=05}
\end{figure}

\begin{figure}[htp]
  \centering
  \includegraphics[width=0.45\textwidth]{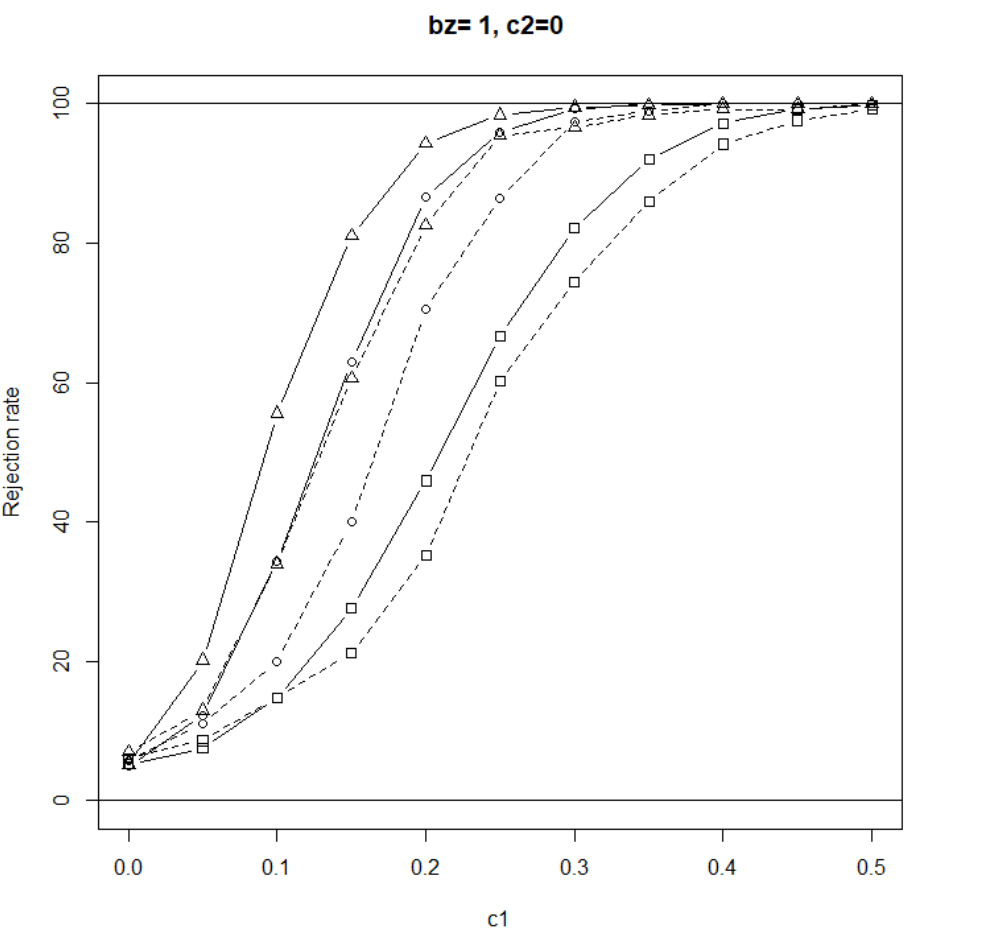}
    \includegraphics[width=0.45\textwidth]{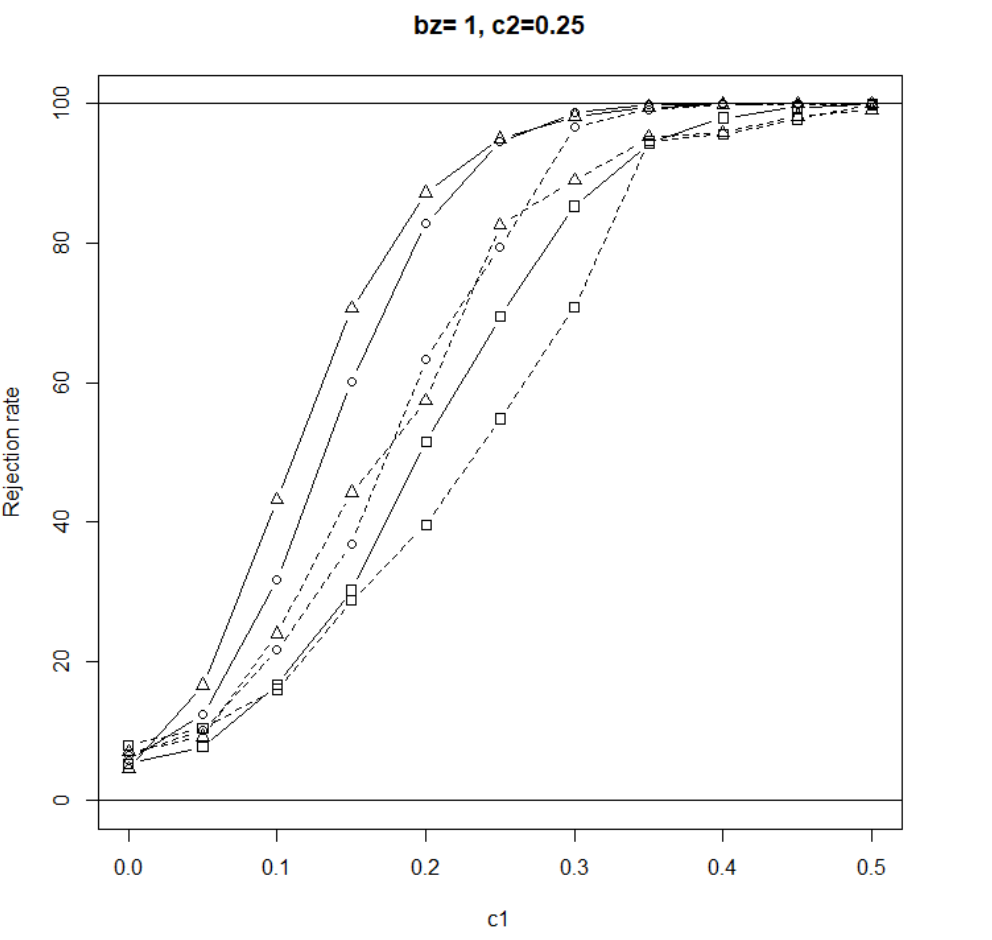}  \\
  \includegraphics[width=0.45\textwidth]{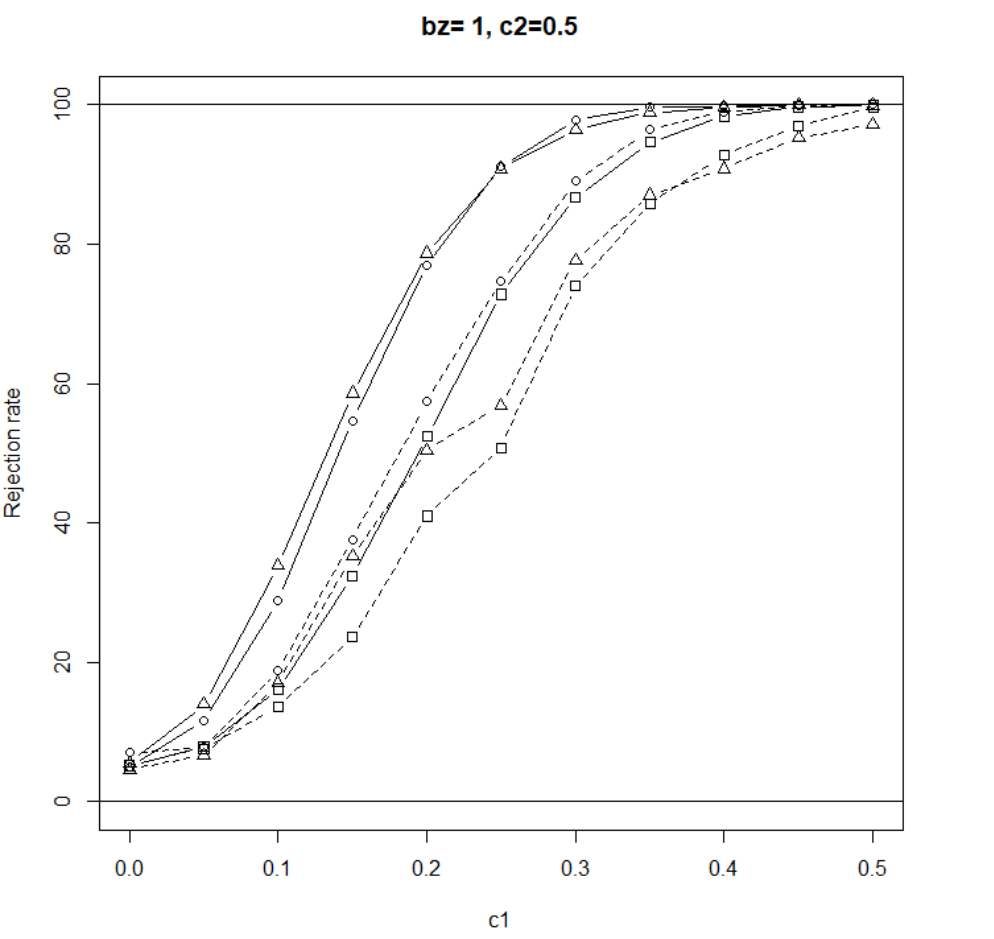}
    \includegraphics[width=0.45\textwidth]{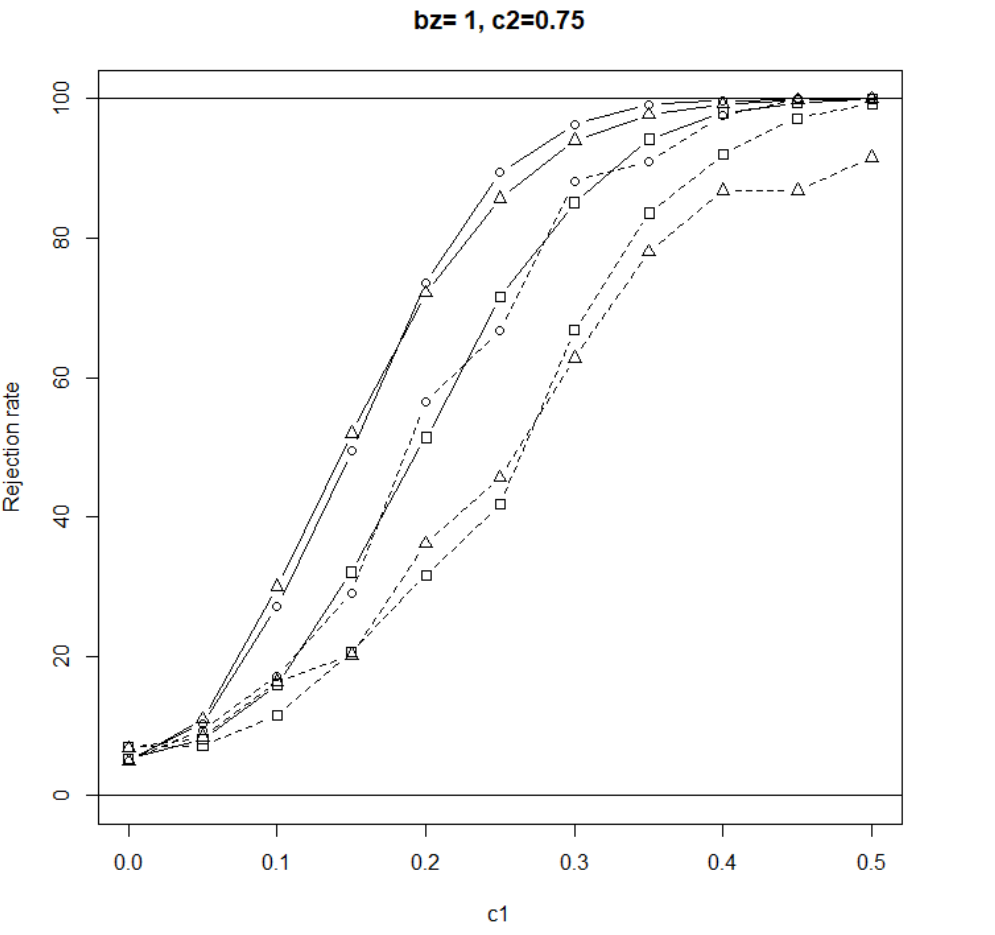}
  \caption{Plots of rejection rates when $b_z=1$ for
 the S2 test (solid lines) and the DUAN test  (dotted lines):
  $w(y) = y$ (circles);
$w(y) = 0.5y^2$ (triangles);
$w(y) = 2.5 I(y>1)$ (squares).
  }\label{power-bz=1}
\end{figure}

For better comparison, we display the power (versus $c_1$) lines
of the S2 and DUAN tests in Figs.~\ref{power-bz=05} and \ref{power-bz=1}, corresponding to $b_z=0.5$ and 1, respectively.
The power lines of the S1 test coincide with those of the S2 test and  hence
are omitted.
It is clear that the power lines of the S2 test  always  lie
above those of the DUAN test, or S2 is uniformly more powerful,
 except for two scenarios
where $b_z=0.5, c_2=0$, and $w(y) = 0.4y^2$ or $2.5I(y>1)$.
In the two exceptional cases, compared with DUAN,
S2 is more powerful  for small  $c_1$
and becomes less powerful for large $c_1$.
As $c_1$ quantifies the departure of the
 true missingness mechanism from the null hypothesis,
a possible explanation for this phenomenon is that
the score test is usually most powerful for `local'  alternatives,
but may be suboptimal when the alternative is not very local.

\begin{table}[htp]
\small
\centering
\caption{Empirical rejection rates (\%) of
the S1, S2, and DUAN  tests
based on 5000 simulated samples from  Example \ref{ex-duan}.
The sample size is $n = 1000$ and $b_z = 0.5$.
}\label{tab-ex-duan-1000-weak}
\tabcolsep  4pt
\renewcommand{\arraystretch}{1}
\begin{tabular}{ccc| ccccccccccc}
  \hline
  &&&  \multicolumn{10}{|c}{$c_1$}  \\
  $w(y)$ & $c_2$& Test & 0 &0.05&0.1&0.15&0.2&0.25&0.3&0.35&0.4&0.45&0.5 \\\hline
  $y$          &  0 & S1   &   5.0&6.5&12.5&20.8&32.3&46.3&59.9&70.7&81.3&87.9&92.7\\
               &    & S2   &   5.0&6.5&12.5&20.8&32.4&46.3&59.9&70.8&81.3&87.9&92.8\\
               &    & DUAN &  5.2&7.6&12.4&17.0&27.8&37.0&55.3&60.8&72.3&84.2&88.2 \\  \cline{2-14}
               &0.25& S1   &  5.2&6.4&12.5&19.8&30.2&43.0&56.2&68.1&78.4&86.0&92.0 \\
               &    & S2   &  5.1&6.4&12.4&19.8&30.4&43.0&56.5&68.2&78.6&86.2&92.1  \\
               &    & DUAN &  7.0&7.2&11.4&16.0&23.0&36.4&49.8&52.6&70.7&76.4&84.4\\   \cline{2-14}
               &0.50& S1   &  5.0&6.6&11.1&18.6&30.1&43.5&55.9&68.9&79.8&87.1&92.4 \\
               &    & S2   &  5.0&6.6&11.0&18.6&30.0&43.8&56.2&69.1&80.2&87.3&92.7 \\
               &    & DUAN &  5.8&7.6&10.4&14.0&24.4&29.0&38.6&57.8&66.8&78.2&85.6\\   \cline{2-14}
               &0.75& S1   &  5.2&6.6&11.9&20.3&31.5&44.4&58.4&70.3&81.3&87.9&92.8 \\
               &    & S2   &  5.1&6.4&11.8&20.0&31.2&44.8&59.0&70.7&81.4&88.1&93.1  \\
               &    & DUAN &  4.4&6.4& 9.4&11.6&20.6&28.6&39.4&49.0&54.6&68.6&82.4\\
      \hline
$0.4y^2$       &  0 & S1   & 4.9&8.5&17.3&27.1&35.9&45.4&54.7&60.5&65.3&71.4&76.1\\
               &    & S2   & 4.9&8.5&17.3&27.1&35.8&45.4&54.7&60.3&65.3&71.3&75.9\\
               &    & DUAN & 5.0&7.8&11.4&19.0&27.8&35.8&54.8&63.9&65.2&69.6&72.6\\   \cline{2-14}
               &0.25& S1   & 4.9&6.4&11.4&17.6&23.0&27.6&36.5&41.4&48.9&56.3&62.0\\
               &    & S2   & 4.9&6.4&11.4&17.4&22.7&27.2&35.9&41.0&48.3&55.3&61.1\\
               &    & DUAN & 4.6&6.0& 9.6&13.6&19.8&29.0&27.9&35.3&45.6&45.6&55.9\\   \cline{2-14}
               &0.50& S1   & 5.1&6.2&9.3&12.2&16.5&22.8&28.1&34.4&42.3&49.16&55.6\\
               &    & S2   & 5.0&6.2&9.1&11.9&16.2&22.0&27.0&33.2&40.7&47.5&53.4\\
               &    & DUAN & 6.4&6.0& 8.6&10.8&16.2&19.0&23.6&33.4&30.5&41.6&39.9\\   \cline{2-14}
               &0.75& S1   & 5.0&6.3&8.7&11.4&16.1&22.8&29.5&37.2&46.8&54.2&63.4\\
               &    & S2   & 5.0&6.1&8.0&10.7&15.0&21.2&27.5&34.8&44.4&51.6&60.5\\
               &    & DUAN & 5.4&6.4& 5.2& 8.4&11.4&19.2&15.6&22.6&33.4&31.0&32.2\\
      \hline
$2.5I(y>1)$    &  0 & S1   &5.2&5.7&7.0&9.9&14.1&20.9&26.7&36.3&45.8&54.4&64.3\\
               &    & S2   &5.2&5.7&7.0&9.9&14.1&20.9&26.7&36.3&45.9&54.5&64.3\\
               &    & DUAN &5.2&7.6& 8.6& 9.6&16.4&21.4&22.8&33.0&45.8&56.8&72.0\\   \cline{2-14}
               &0.25& S1   &5.0&5.4&8.2&11.8&17.1&23.6&33.0&43.7&54.6&65.4&75.8\\
               &    & S2   &5.0&5.4&8.1&11.8&17.0&23.6&33.0&43.7&54.3&65.2&75.6 \\
               &    & DUAN &6.0&9.2& 7.8&11.2&11.4&16.4&35.0&37.0&52.5&60.4&65.4\\   \cline{2-14}
               &0.50& S1   &4.6&6.1&8.9&13.2&20.8&28.8&39.9&54.1&65.9&76.1&85.7 \\
               &    & S2   &4.6&6.1&8.8&13.1&20.5&28.5&39.6&53.7&65.4&75.8&85.4 \\
               &    & DUAN &6.4&6.6& 6.4&11.6&16.2&23.6&30.8&40.2&46.2&56.2&68.2\\   \cline{2-14}
               &0.75& S1   &4.8&6.5&10.1&16.3&24.8&34.1&48.7&62.2&72.6&83.4&90.4 \\
               &    & S2   &4.9&6.4&9.9&16.1&24.5&33.8&48.1&61.6&72.3&82.8&90.2 \\
               &    & DUAN &5.6&5.4& 8.0&10.0&13.6&18.8&36.6&37.2&49.4&53.4&62.8\\
      \hline
\end{tabular}
\end{table}

\begin{table}[htp]
\small
\centering
\caption{
Empirical rejection rates (\%) of
the S1, S2, and DUAN  tests
based on 5000 simulated samples from  Example \ref{ex-duan}.
The sample size is $n = 1000$ and $b_z = 1$.
}\label{tab-ex-duan-1000-strong}
\tabcolsep  4pt
\renewcommand{\arraystretch}{1}
\begin{tabular}{ccc|ccccccccccc}
  \hline
   &&& \multicolumn{10}{|c}{$c_1$}  \\
  $w(y)$ & $c_2$& Test & 0 &0.05&0.1&0.15&0.2&0.25&0.3&0.35&0.4&0.45&0.5 \\\hline
  $y$ &  0 & S1   &5.0&12.2&34.4&63.0&86.6&95.9&99.3&99.9&100.0&100.0&100.0 \\
      &    & S2   &5.0&12.2&34.4&63.0&86.6&95.9&99.3&99.9&100.0&100.0&100.0 \\
      &    & DUAN &5.8& 11.0& 20.0& 40.0& 70.6& 86.4& 97.4& 99.0& 100.0& 100.0& 100.0\\   \cline{2-14}
      &0.25& S1   &5.7&12.4&31.8&60.2&82.9&94.5&98.7&99.8&100.0&100.0&100.0 \\
      &    & S2   &5.7&12.4&31.8&60.2&82.9&94.5&98.7&99.8&100.0&100.0&100.0 \\
      &    & DUAN &6.6& 10.2& 21.6& 36.8& 63.4& 79.4& 96.6& 99.2&  99.8&  99.8& 100.0\\   \cline{2-14}
      &0.50& S1   &4.9&11.6&28.9&54.8&77.0&91.2&97.8&99.6&99.8&100.0&100.0 \\
      &    & S2   &5.0&11.6&28.8&54.7&77.0&91.2&97.8&99.6&99.8&100.0&100.0 \\
      &    & DUAN &7.0&  7.8& 18.8& 37.6& 57.4& 74.6& 89.0& 96.4&  99.0&  99.8& 100.0\\   \cline{2-14}
      &0.75& S1   &5.2&10.4&27.2&49.4&73.6&89.5&96.2&99.1&99.7&100.0&100.0 \\
      &    & S2   &5.1&10.3&27.2&49.5&73.5&89.4&96.3&99.2&99.7&100.0&100.0 \\
      &    & DUAN &5.0&  9.4& 17.2& 29.0& 56.6& 66.8& 88.2& 91.0&  97.6&  99.8& 100.0\\
      \hline
  $0.4y^2$
      &  0 & S1   & 5.2& 20.2& 55.6& 81.1& 94.3& 98.4& 99.5& 99.8& 100.0& 100.0& 100.0\\
      &    & S2   & 5.2& 20.2& 55.6& 81.1& 94.3& 98.4& 99.5& 99.8& 100.0& 100.0& 100.0\\
      &    & DUAN &7.0&13.0&34.0&60.6&82.6&95.4&96.6& 98.4& 99.2& 99.2&100.0\\\cline{2-14}
      &0.25& S1   &4.5& 16.6& 43.2& 70.7& 87.3& 95.1& 98.2& 99.5& 99.9& 100.0& 100.0 \\
      &    & S2   &4.5& 16.6& 43.2& 70.6& 87.2& 95.0& 98.2& 99.5& 99.9& 100.0& 100.0 \\
      &    & DUAN &7.0& 9.2&24.0&44.2&57.4&82.6&89.0& 95.2& 95.8& 98.2& 99.0\\\cline{2-14}
      &0.50& S1   &5.6& 14.1& 33.9& 58.8& 78.8& 90.8& 96.5& 98.9& 99.6& 100.0& 99.9 \\
      &    & S2   &5.5& 14.0& 33.9& 58.6& 78.7& 90.8& 96.4& 98.9& 99.6& 100.0& 99.9 \\
      &    & DUAN &4.6&06.6&17.0&35.2&50.4&56.8&77.6& 87.0& 90.8& 95.2& 97.2\\\cline{2-14}
      &0.75& S1   &5.0& 11.1& 30.2& 52.2& 72.3& 86.1& 94.1& 97.8& 99.3& 99.8& 100.0 \\
      &    & S2   &4.9& 11.0& 30.0& 52.0& 72.1& 85.7& 94.0& 97.7& 99.2& 99.8& 100.0 \\
      &    & DUAN &6.8& 8.4&16.4&20.2&36.2&45.6&62.8& 78.0& 86.8& 86.8& 91.6\\
      \hline
  $2.5I(y>1)$
      &  0 & S1   &5.3& 7.5& 14.7& 27.7& 45.9& 66.7& 82.1& 92.0& 97.2& 99.2& 99.8\\
      &    & S2   &5.3& 7.5& 14.7& 27.7& 45.9& 66.7& 82.1& 92.0& 97.2& 99.2& 99.8 \\
      &    & DUAN &6.0& 8.6&14.8&21.2&35.2&60.2&74.4&86.0&94.2&97.6& 99.2\\      \cline{2-14}
      &0.25& S1   &5.3& 7.7& 16.6& 30.2& 51.5& 69.6& 85.4& 94.3& 98.0& 99.5& 99.9 \\
      &    & S2   &5.3& 7.7& 16.6& 30.2& 51.5& 69.5& 85.3& 94.3& 98.0& 99.5& 99.9 \\
      &    & DUAN &8.0&10.4&16.0&28.8&39.6&54.8&70.8&94.6&95.6&97.8& 99.8\\     \cline{2-14}
      &0.50& S1   &5.2& 7.9& 16.2& 32.3& 52.4& 72.7& 86.8& 94.6& 98.3& 99.7& 99.9 \\
      &    & S2   &5.2& 7.9& 16.2& 32.3& 52.4& 72.8& 86.7& 94.6& 98.3& 99.7& 99.9 \\
      &    & DUAN &5.2& 7.6&13.6&23.6&41.0&50.8&74.0&85.8&92.8&97.0& 99.6\\      \cline{2-14}
      &0.75& S1   &5.4& 8.1& 16.0& 32.2& 51.5& 71.5& 85.0& 94.1& 98.0& 99.4& 99.9 \\
      &    & S2   &5.3& 8.1& 16.0& 32.1& 51.4& 71.5& 85.1& 94.2& 98.0& 99.4& 99.9 \\
      &    & DUAN &7.0& 7.2&11.6&20.6&31.6&41.8&66.8&83.6&92.0&97.2& 99.2\\
      \hline

\end{tabular}
\end{table}

\begin{example}\label{ex-no-instrument}
Let $X$ denote a univariate  covariate   following $\mathbb{N}(0,1)$.
The conditional distribution of  $Y$ given $X=x$
is $\mathbb{N}(\xi_1 x+\xi_2x^2, \exp(\xi_3+\xi_4 x))$, with $\xi=(\xi_1, \ldots, x_4)=(-1, 1, 0.5, 0)$
or $(1, 1, 0.5, 1)$.
The missingness indicator   $D$ of  $Y$ conditional on $(Y=y, X=x)$
follows a logistic model,
$\pr(D=1|x, y) = \pi(\beta_0+\beta_1 x + \gamma y)$.
We consider six choices of  $(\beta_0, \beta_1)$,  namely
  $(0.85, 0)$, $(0.6, 0.25)$, $(0.4, 0.5)$ in the case of $\xi=(-1,1,0.5,0)$
  and $(0.85, 0)$, $(0.7, 0.25)$ $(0.5,0.5)$ in the case of $\xi = (1,1,0.5,1)$.
These settings  are chosen such that   the  missingness rates are about $20\%$--$30\%$.
The parameter $\gamma$ is set to 0, 0.05, \ldots, and 0.25, respectively.
\end{example}

Example~\ref{ex-no-instrument} is designed
to represent the case where no instrument is present,
and therefore the DUAN test is not applicable.
The choices of $\xi_4=0$ and 1 correspond
to  a homogeneous variance and a heterogeneous variance, respectively.
We take $f(y|x, \xi)$  and $\mu(x, \theta)$
in the constructions of  S1 and S2
to be the density functions of $\mathbb{N}(\xi_1 x+\xi_2x^2, \exp(\xi_3+\xi_4 x))$
and  $\theta_1 x+\theta_2 x^2$, respectively,  where $\theta=(\theta_1, \theta_2)^\T$.
Table  \ref{tab-ex-no-instrument} presents the simulated rejection rates of the
S1 and S2 tests when data are generated
from Example~\ref{ex-no-instrument}
and the sample size  $n=1000$.
The results corresponding to $\gamma=0$ are  type I errors,
and the type I errors of both S1 and S2 are under control.
As $\gamma$ increases from 0 to 0.25,
both tests have desirable and increasing powers whether the variance is  homogeneous or heterogeneous.
Again, the results for both tests are all nearly equal to each other in all cases.
As S1 requires stronger model assumptions, it may be more risky
for model mis-specification than S2.  Hence, we would recommend using S2 rather than S1
for testing whether the missingness mechanism is ignorable missing or nonignorable missing.

\begin{table}[htp]
\centering
\caption{
Empirical rejection rates (\%) of
the S1 and S2   tests
based on 5000 simulated samples of size $n = 1000$
from  Example \ref{ex-no-instrument}.
} \label{tab-ex-no-instrument}
\begin{center}
\tabcolsep 12pt
\renewcommand{\arraystretch}{1}
\begin{tabular}{ccc|cccccc}
  \hline
         &  && \multicolumn{6}{|c}{  $\gamma$}  \\ 
$\xi$             & $\beta_1$& Test & 0 & 0.05 & 0.1 & 0.15 & 0.2 & 0.25 \\  \hline
$ (-1,1,0.5,0)$  &0   &S1&  4.7& 17.4& 50.9& 81.7& 95.6& 99.2\\
                 &    &S2&  4.8& 17.3& 50.8& 81.7& 95.5& 99.2\\   \cline{2-9}
                 &0.25&S1&  4.9& 17.4& 50.6& 81.7& 96.4& 99.4\\
                 &    &S2&  4.9& 17.2& 50.4& 81.6& 96.4& 99.4\\   \cline{2-9}
                 &0.5 &S1&  5.4& 16.9& 47.6& 79.3& 94.8& 99.1\\
                 &    &S2&  5.4& 16.8& 47.3& 79.0& 94.7& 99.1\\   \cline{2-9}
                 &1   &S1&  4.7& 13.0& 35.8& 66.0& 86.3& 97.3\\
                 &    &S2&  5.1& 12.8& 35.3& 65.4& 86.1& 97.1\\   \hline
$ (1,1,0.5,1)$   & 0  &S1&  4.6& 14.4& 37.4& 60.9& 77.7& 87.4\\
                 &    &S2&  5.0& 13.0& 36.3& 60.4& 77.5& 88.4\\ \cline{2-9}
                 &0.25&S1&  5.2& 13.6& 35.1& 57.4& 75.9& 86.6\\
                 &    &S2&  5.3& 13.4& 34.7& 57.6& 76.5& 87.4\\ \cline{2-9}
                 &0.5 &S1&  4.7& 14.2& 33.6& 55.3& 73.7& 86.4\\
                 &    &S2&  4.7& 14.1& 34.3& 56.2& 74.6& 87.0\\ \cline{2-9}
                 & 1  &S1&  4.7& 11.0& 28.1& 47.2& 66.0& 79.9\\
                 &    &S2&  4.6& 11.1& 27.5& 46.8& 65.2& 79.4\\ \hline
\end{tabular}
\end{center}
\end{table}

\section{ Application to human immunodeficiency virus data }
\label{s:data}
 For illustration, we analyse   human immunodeficiency virus (HIV) data from
 AIDS Clinical Trials Group Protocol 175  \citep{h96, h19, l20}.
These data are available from the {\tt R} package {\tt speff2trial}
and consist of various measurements of $n =2139$ HIV-infected patients.
The patients were randomly divided into four arms according
 to the regimen of treatment they received:
 (I) zidovudine monotherapy, (II) zidovudine + didanosine,
 (III) zidovudine + zalcitabine and (IV) didanosine monotherapy.
Important measurements from the patients
 include
 CD4 cell count at baseline (cd40),
 CD4 cell count at 20$\pm$5 weeks (cd420),
 CD4 cell count at 96$\pm$5 weeks (cd496),
 CD8 cell count at 20$\pm$5 weeks (cd820)
 and arm number (arms).
The effectiveness of a HIV treatment can be assessed
 by monitoring the CD4 cell counts of HIV-positive patients:
 an increase in such counts is an indication of improvement
in the patients' health.
The typical problem of  interest is to estimate the mean of the CD4 cell counts
 in each arm after the patients were treated for about 96 weeks.

We take cd496 as a response variable $Y$
 and we take   cd40, cd420 and cd820
 as covariates $X1$, $X2$ and $X3$, respectively.
 Owing to the end of the trial or loss to follow-up,
 39.66\% of the patients' responses were missing.
The analyses of \cite{h96} and \cite{h19} were based on
the MAR assumption,
whereas   \cite{l20} and \cite{z20} assumed that the response was
nonignorable missing.
As   correctly specifying the underlying missingness mechanism is
crucial to  validation of the subsequent inference,
it is also necessary to formally check whether
the missingness is MAR or not.

Let $\mathbf{X} = (1, X_1,X_2,X_3)$.  We
choose $f(y|\mathbf{x}, \xi)$ to be the normal density
with mean $\mu(\mathbf{x},\xi)=\xi_1 + \xi_2 x_1 + \xi_3 x_2 +\xi_4 x_3 + \xi_5 x_2^2$
and variance $\sigma(\mathbf{x},\xi)=\xi_6$, where $\xi=(\xi_1, \ldots, \xi_6)^\T$.
After explorative analysis, we find that
among the three covariates,
only  $x_2$ is significantly correlated with
the missingness indicator.
We assume that the missingness indicator in each regimen of treatment
follows a linear logistic regression  model with covariates $x_2$ and $y$ only.
As no instrumental variable is present  in this application, we apply
only the proposed two scores to test
whether the missingness of cd496 depends on itself.

The $p$-values of the proposed two score tests are reported in Table~\ref{HIV-data}.
None of the results are significant at the 5\%  level.
In other words, they all support the MAR  mechanism in the four regimens.
At the same time, both the tests have very close $p$-values.
Table~\ref{HIV-data} also presents their $p$-values
if  we remove the covariate $x_2$ from the propensity score model.
The $p$-values for regimens II and IV are seemingly unchanged and insignificant.
However, those for regimens I and III become much less, and much smaller than the 5\% significance level.
These results indicate that  the MNAR  mechanism seems
more reasonable than MAR if the propensity score  depends potentially on  $y$.
A possible explanation for the insignificant result in the presence of $x_2$
is that $x_2$ and $y$
stand for  CD4 cell counts at $20\pm5$ weeks and   at $96\pm5$ weeks, respectively,
and they are highly correlated.

\begin{table}[htp]
\centering
\caption{P-values of the proposed score tests S1 and S2
under the four  regimens of treatment based on the HIV data
\label{HIV-data}
}
\begin{tabular}{ c cccc }
  \hline
  Treatment regimen     & I            & II        & III          & IV  \\\hline
    \multicolumn{5}{c}{ $X_2$  appears in the logistic model }  \\
  S1& 0.3263 &0.3730 & 0.4490 &  0.2081 \\\hline
  S2& 0.1291 & 0.3388 &0.3730 &  0.1584   \\\hline
   \multicolumn{5}{c}{$X_2$ does not appear in the logistic model}  \\
  S1&  0.0065 &0.3731   &0.0104   & 0.2081    \\\hline
  S2&  0.0003 &0.3389   &0.0006   & 0.1584     \\\hline
\end{tabular}
\end{table}

\section{Discussion}
\label{s:discuss}

 Valid data analyses of missing data rely on a correctly specified missingness mechanism.
The problem of  testing whether the missingness mechanism is MCAR or MAR is relatively easy to solve and has been extensively studied. However,  it is much more challenging
to test whether the  mechanism is   MAR  or not,
because parameters may no longer be identifiable.
We avoid this thorny issue by using a score test,
which is constructed under the null hypothesis, namely
the MAR mechanism.   The underlying parameters are usually identifiable
based on MAR data. This is one of the nice properties of a score test
 \citep{r05}. A score test is also invariant under transformation of  parameters.
Transformation of parameters may simplify parameter estimation without affecting the value of the statistic.
We derive two score tests, S1 and S2,  when
the conditional density of $Y$ given $X$ is modelled by
a completely parametric model and a semiparametric location model, respectively.
Our numerical results indicate that
these tests generally have nearly the same performance (type I error and power), but
S2 is preferable because it requires weaker model assumptions.

\section*{ ACKNOWLEDGEMENTS}


 This research was supported by the National Natural Science Foundation of China (11771144),
the State Key Program of the National Natural Science Foundation of China (71931004 and 32030063),
the Development Fund for Shanghai Talents, and the 111 Project (B14019).
Drs Lu  and Liu   are the corresponding authors.

\section*{Supplementary Materials}

\label{SM}

The proofs of Theorems \ref{score1-normality}--\ref{score-local-power} and more simulation results are available with
this paper at the Supplementary Materials.

\end{document}